\documentclass[a4paper,11pt]{article}
\pdfoutput=1 

\usepackage{jcappub} 

\usepackage[T1]{fontenc} 

\usepackage{array}
\usepackage{multirow}
\usepackage{multicol}
\usepackage[normalem]{ulem}

\usepackage[multiple]{footmisc} 
\usepackage{hyperref}

\title{\boldmath Nonlinear structure formation in Bound Dark Energy}

\author[a,b,1]{Erick Almaraz,\note{Corresponding author.}}
\author[c]{Baojiu Li,}
\author[a]{and Axel de la Macorra}

\affiliation[a]{Instituto de F\'isica, Universidad Nacional Aut\'onoma de M\'exico,\\04510, Ciudad de M\'exico, Mexico}
\affiliation[b]{African Institute for Mathematical Sciences,\\6 Melrose Road, Muizenberg, 7945, South Africa}
\affiliation[c]{Institute for Computational Cosmology, Ogden Centre for Fundamental Physics\\Department of Physics, University of Durham, Science Laboratories,\\South Road, Durham, DH1 3LE United Kingdom}

\emailAdd{ealmaraz@estudiantes.fisica.unam.mx}
\emailAdd{baojiu.li@durham.ac.uk}
\emailAdd{macorra@fisica.unam.mx}

\abstract{We study nonlinear structure formation in the Bound Dark Energy model (BDE), where dark energy (DE) corresponds to a light scalar meson particle $\phi$ dynamically formed at a condensation energy scale $\Lambda_c$. The evolution of this dark-energy meson is determined by the potential $V(\phi)=\Lambda_c^{4+2/3}\phi^{-2/3}$, with a distinguishing phenomenology from other quintessence scenarios. Particularly, the expansion rate of the universe is affected not only at late times, but also when the condensation of $\phi$ occurs, which in linear theory leads to an enhancement (with respect to standard $\Lambda$CDM) of matter perturbations on small scales. We study how much of this signature is still present at late times as well as the properties of dark matter halos in the nonlinear regime through N-body simulations. Our results show that nonlinear corrections  wash out this feature from the matter power spectrum even before DE becomes dominant. There is, however, a small but clear suppression of the BDE spectrum of $2\%$ today on the largest scales due to the distinct late-time dynamics of DE. The differences on the clustering power between BDE and $\Lambda$CDM are reflected in the halo mass function, where small halos are more abundant in BDE as opposed to large heavy structures, whose formation is delayed because of the expansion history of the universe. This result is well captured by the semi-analytical Sheth-Tormen formula. However, despite these differences, the halo concentration parameter is essentially the same in both models, which suggest that clustering inside the halos decouple from the general expansion once the halos form.
}

\begin{document}
\maketitle
\flushbottom

\section{Introduction}\label{sec:introduction}
Research done over the last twenty years has firmly established that the universe is currently expanding at an accelerating rate \cite{Weinberg13_DE,Mortonson13_DE,Huterer17_DE}. Assuming that General Relativity still provides an accurate description of gravity on cosmological scales, the late-time cosmic acceleration can be interpreted as the dynamical effect on the motion of galaxies and structures because of the pressure exerted by dark energy (DE), which in the standard concordance model ($\Lambda$CDM) is fully characterized by a cosmological constant, $\Lambda$. The cosmic abundance of DE is constrained to about 70\% of the energy content of the universe at present time, while 26\% consists of Cold Dark Matter (CDM) and the remaining 4\% is left to the Standard Model (SM) particles \cite{PlanckCP15_CMB}. Despite the success of the concordance model in providing a simple theoretical framework to account for observations, the physical nature of DE is still a mystery. The main problem in attributing DE to a cosmological constant is the disturbing discrepancy between the theoretically predicted estimations of $\Lambda$ and its observed value, an issue that is commonly referred to as \textsl{the fine-tuning problem} \cite{Weinberg89_CC,Martin12_CC,Sola13_CC}. The inability of standard physics to explain this discrepancy impels the quest of alternative DE candidates, such as quintessence \cite{Copeland06_QCDM,Tsujikawa13_QCDM}, modified gravity theories \cite{Clifton12_MG}, and other theoretical scenarios \cite{Bamba12_DE,Joyce15_DE}.

The progress made in observational cosmology since the discovery of the cosmic acceleration  now allows us to assess the viability of these alternative scenarios as well as to look for deviations with respect to $\Lambda$CDM that may be detectable in the future. Such departures from standard $\Lambda$CDM may be looked in the Large Scale Structure (LSS) of the universe, where DE is expected to leave imprints by its effects on the expansion history and the presence of DE perturbations \cite{Linder03_DE, Mainini03_DE,Percival05_DE,Mehrabi15_DE}. In recent times N-body simulations have become a powerful tool to explore the impact of DE on the distribution of matter across the universe, the formation of cosmic structures and their dynamical properties \cite{Kuhlen12_nbody,Baldi12_DEsimulations}. These simulations serve as numerical laboratories to study physical processes on scales where linear perturbation theory is no longer valid. The increase of computational power and the development of more efficient algorithms make it now possible to run large N-body simulations within a reasonable period of time and consequently it paves the way for a systematic study of different DE scenarios \cite{Klypin03_DEsimulations,Dolag04_DEsimulations,Lokas04_DEsimulations,Francis07_DEsimulations,Casarini09_DEsimulations,Grossi09_DEsimulations,Alimi10_QCDM,Courtin11_HMF}.

In this paper we study structure formation in the Bound Dark Energy (BDE) model \cite{Macorra01_BDE,Macorra02_BDE,Macorra03_BDE,Macorra05_BDE,Macorra18_BDE,Almaraz19_BDE}. BDE is a quintessence theory aiming to explain the nature of DE through a natural extension of the SM. Inspired by supersymmetry and unification schemes, the model introduces a hidden dark gauge group (DG) of light particles coupled with the SM sector only through gravity below the unification scale. At high energies the gauge coupling of the DG is weak and the energy density of these light particles dilutes as radiation. When the temperature drops off and a critical energy density scale $\Lambda_c$ is reached, the gauge coupling of the DG becomes strong and now these particles condense into composite states. DE is the lightest composite state corresponding to a scalar meson $\phi$ whose dynamical evolution is determined by an inverse power-law potential (IPL) $V(\phi)=\Lambda_c^{4+2/3}\phi^{-2/3}$. The condensation energy scale $\Lambda_c$, the exponent $\alpha=2/3$ of the potential and the epoch $a_c$ where the condensation of the scalar meson $\phi$ occurs are not free cosmological parameters, but they are derived quantities determined by the properties of the DG. Moreover, the initial conditions of the BDE field at $a_c$ are also determined by the symmetry breaking scale $\Lambda_c$ of the DG and therefore the amount of DE at any time can be straightforwardly predicted from the solution of the background equations. Unlike other quintessence theories, in this scenario the scalar field representing DE is not a fundamental entity in nature, but it results from the interaction between particles that have the same primitive status as the other fundamental particles of the SM. Therefore, even tough the BDE potential turns out to be a particular case of the well-known Ratra-Peebles IPL potential $V(\phi)=M^{4+\alpha}\phi^{-\alpha}$ \cite{Peebles88_QCDM,Ratra88_QCDM,Wetterich88_QCDM}, BDE entails a complete different cosmological scenario, where the dynamical evolution of DE, its effects on the expansion history of the universe, and the implications of the model are not the same.  

We developed these ideas and presented the constraints on the BDE model in \cite{Macorra18_BDE,Almaraz19_BDE}. Our analysis focused on the predictions arising from the background dynamics and the linear perturbation theory. BDE fits well the data, particularly Baryon Acoustic Oscillations (BAO) measurements, where we found a systematic better fit than standard $\Lambda$CDM  leading to interesting tensions that might be useful to discriminate between these two scenarios. Additionally, we also found interesting imprints on the matter power spectrum, where the condensation of the scalar BDE meson leads to an enhancement of power on small scales, while the power on large scales is suppressed by the late-time dynamics of DE. In view of the importance of LSS studies on DE research in the forthcoming years, here we extend our previous work to investigate the clustering of matter in the BDE model in the nonlinear regime.

This paper is organized as follows. In section \ref{sec:bde_model} we make a general review of BDE. This review comprehends the theoretical insights and the basic equations of the model, the discussion of some interesting deviations with respect to $\Lambda$CDM, and the description of the DE dynamics and its effect on the cosmic expansion. In section \ref{sec:bde_linear_quasilinear} we analyze the evolution of matter perturbations first in the linear theory, and then we study the transition to the nonlinear regime according to the spherical collapse model. We study the nonlinear regime in section \ref{sec:bde_nonlinear}. We describe the setup of our N-body simulations and present our results on the matter power spectrum, the halo mass function, and the concentration parameter. Finally, we summarize our findings and state our conclusions in section \ref{conclusions}. Here we adopt the usual notational conventions found in the literature. Particularly, $a$ denote the scale factor of the universe, which is related with the cosmological redshift $z$ by $1+z=1/a$ for a flat Friedmann-Lema\^itre-Robertson-Walker (FLRW) metric, $G$ is the gravitational constant, and overdots denote cosmic-time derivatives.

\section{The Bound Dark Energy model}\label{sec:bde_model}
Our Dark Energy model \cite{Macorra01_BDE,Macorra02_BDE,Macorra03_BDE,Macorra05_BDE,Macorra18_BDE,Almaraz19_BDE} introduces a supersymmetric $SU(N_c)$ gauge group of light particles with $N_c=3$ colors and $N_f=6$ flavors in the fundamental representation. Since this Dark Group is postulated as an extra ingredient of the constituents of matter, neither $N_c$ nor $N_f$ represent cosmological free quantities, but they  are input parameters whose fundamental status is the same as the other input parameters of the SM. The DG is unified with the SM sector at the unification scale $\Lambda_\mathrm{GUT}\approx 10^{16}\mathrm{GeV}$, below which they interact only via gravity. At high energies the gauge coupling of the DG is weak, so the light particles of the group are asymptotically free and the DG contributes to the total amount of radiation of the universe. The gauge coupling of the DG evolves over time growing at low energies as the universe expands and cools down. When a critical energy scale $\Lambda_c$ is reached at a scale factor $a_c$ (related each other by $a_c\Lambda_c/\textrm{eV}=1.0939\times 10^{-4}$ in a 3 massless neutrino species scenario), the coupling becomes strong and the particles of the DG bind together forming composite states. The lightest formed state corresponds to a scalar meson described by a real scalar field with an IPL self-interaction term 
\begin{equation}\label{eq:bde_V}
V(\phi)=\Lambda_c^{4+2/3}\phi^{-2/3}.
\end{equation}
This meson represents DE and it is precisely the aforementioned binding mechanism where the name BDE is given to our dark energy model.

After the particles of the DG condense into BDE, the extra relativistic degrees of freedom of the DG vanish and the cosmological evolution of the scalar field is analytically described by the canonical quintessence formalism \cite{Copeland06_QCDM,Tsujikawa13_QCDM}. The evolution of the scalar field at the homogeneous-background level is governed by the Klein-Gordon equation
\begin{equation} \label{eq:Klein-Gordon} 
\ddot{\phi}+3H\dot{\phi}+\frac{dV}{d\phi} = 0,
\end{equation}
with a frictional term depending on the expansion rate of the universe ($H \equiv \dot{a}/a$), and the steepness of the scalar potential $dV/d\phi$ acting as a driving force. The energy density and the pressure of the field are given by
\begin{equation}\label{eq:bde_rhophi_Pphi}
\rho_\mathrm{BDE}= \frac{1}{2}\dot{\phi}^2+V, \hspace{1.5cm} P_\mathrm{BDE}=\frac{1}{2}\dot{\phi}^2-V,
\end{equation}
leading to a time-varying equation of state (EoS)
\begin{equation} \label{eq:bde_wphi} 
w_\mathrm{BDE} \equiv \frac{P_\mathrm{BDE}}{\rho_\mathrm{BDE}} =\frac{\frac{1}{2}\dot{\phi}^2-V}{\frac{1}{2}\dot{\phi}^2+V},
\end{equation}
whose value depends on the competition between the kinetic and potential energy of the field. Note that $w_\mathrm{BDE}$ is not allowed to cross to the phantom regime $w<-1$, since $\phi$ is real. Therefore, when comparing with observations the constraints on the EoS will automatically lie outside the phantom region.	This implicit restriction is not present in other DE alternative scenarios, such as scalar field doublets and phenomenological parametrizations of the EoS \cite{Copeland06_QCDM}, where is possible that $w<-1$ and therefore it might happen that these models fit the data just in the phantom region. The main point is that the constraints on the EoS depend strongly on the model being tested as well as on the data used in the fit \cite{PlanckCP15_CMB,PlanckDE15_CMB}.

Although eqs. (\ref{eq:Klein-Gordon})-(\ref{eq:bde_wphi}) coincide with the Ratra-Peebles potential for $\alpha=2/3$, we stress that both models differ in some important features. First, in BDE the scalar field representing DE is not present in the very early universe, but it arises after the particles of the DG condense at relatively late times. Before that, there is an extra amount of radiation which impacts the synthesis of primordial elements. On the other hand, the parameters of the potential $\Lambda_c$ and $\alpha$ as well as other relevant quantities such as $a_c$ are connected with the basic properties of the DG instead of being determined by cosmological observations \cite{Almaraz19_BDE}. In consequence, the evolution of DE in both models is different (see, for example \cite{Alimi10_QCDM}).

We presented the constraints on the BDE model using recent measurements of the CMB temperature anisotropy spectrum \cite{PlanckCP15_CMB}, the distance modulus of type Ia supernovae (SNeIa) \cite{Betoule14_SNeIa} and the BAO distance ratio \cite{Ross15_BAO,Beutler11_BAO,GilMarin16_BAO} in \cite{Macorra18_BDE,Almaraz19_BDE}. Our modified version of the Boltzmann code \texttt{CAMB} \cite{Lewis00_CAMB} used to solve the background and linear perturbation equations, as well as the modified version of \texttt{CosmoMC} \cite{Lewis02_CosmoMC} used to sample the parameter space are publicly available.\footnote{\url{https://github.com/ealmaraz/Bound\_Dark\_Energy\_CAMB}}\footnote{\url{https://github.com/ealmaraz/Bound\_Dark\_Energy\_CosmoMC}} As we mentioned before, $\Lambda_c$ is not a free cosmological parameter, but it can be derived from fundamental physics. However, present uncertainties on the unification scale $\Lambda_\mathrm{GUT}$ and the unified coupling constant $g_\mathrm{GUT}$ lead to a still-inaccurate theoretical prediction of $\Lambda_c^\mathrm{th}=34^{+16}_{-11}\textrm{ eV}$ and consequently we varied $\Lambda_c$ in our MCMC analysis. Nevertheless, the initial conditions of $\phi$ at $a_c$ are not arbitrary, but also depend on $\Lambda_c$ and they don't need to be tuned to retrieve an input value of the DE density today. Given the present content of matter $\Omega_m h^2$ and $\Lambda_c$, the expansion rate today $H_0$ follows directly from the solution of the background equations and therefore the set of hard parameters to sample in BDE is the same size as in $\Lambda$CDM, with $H_0$ replaced by $\Lambda_c$ as a primary parameter.

According to our results, condensation of BDE occurs at $a_c=(2.48 \pm 0.02)\times 10^{-6}$ and the energy scale of condensation is $\Lambda_c = 44.09 \pm 0.28 \textrm{ eV}$, which remarkably lies within the range of our theoretical prediction $\Lambda_c^\mathrm{th}$. The model is in good agreement with observations. When comparing to $\Lambda$CDM we may look into some information criteria, such as the Akaike information criterion (AIC) \cite{Akaike74_Statistics} and the Bayesian information criterion (BIC) \cite{Schwarz78_Statistics} defined respectively as
\begin{equation}\label{eq:AIC}
\mathrm{AIC} = -2\ln \mathcal{L}+2k, \hspace{1.5cm}
\mathrm{BIC}=-2\ln \mathcal{L}+k\ln N,
\end{equation}where $\mathcal{L}$ is the likelihood, $k$ is the number of hard parameters, and $N$ is the number of points used in the fit. Since $k$ and $N$ are the same in both models, any difference in these criteria is solely due to the likelihood \cite{Liddle04_Statistics}. We obtained $\Delta \mathrm{AIC}^{(\mathrm{BDE}-\Lambda\mathrm{CDM})}=\Delta \mathrm{BIC}^{(\mathrm{BDE}-\Lambda\mathrm{CDM})}= -1.2$ favoring BDE for the best fit point. The main difference comes from BAO observations, where BDE outperforms $\Lambda$CDM by $\mathcal{L}_\mathrm{BDE}/\mathcal{L}_{\Lambda\mathrm{CDM}}=e^{-\Delta \chi^2_\mathrm{BAO}/2}=2.1$. Interestingly, BAO and $H_0$ constraints can be combined \cite{Almaraz19_BDE} to lead to tensions with respect to $\Lambda$CDM that can be studied with more detail by upcoming missions.\footnote{\url{https://www.desi.lbl.gov/}}\footnote{\url{https://www.lsst.org/}}

Another interesting feature of the model concerns the effect of the DG on the production of light elements during the Big Bang Nucleosynthesis era (BBN). The presence of the DG represents an extra source of radiation that enhances the expansion rate in the early universe, which results in a higher temperature of decoupling $T_f\sim(H/G_F^2)^{1/5}$ (where $G_F$ is the Fermi constant) between neutrinos and electrons + positrons, before BBN begins. The resulting increased value of $T_f$ enhances the neutron-to-proton ratio $(n/p)=e^{-(m_n-m_p)/T_f}$ (where $m_n$ and $m_p$ is the neutron and proton mass, respectively) which in turn favors the synthesis of helium \cite{Cyburt16_BBN}. Using the interpolation grid from the \texttt{PArthENoPE} code \cite{Pisanti08_Parthenope} available in \texttt{CosmoMC} to estimate the abundance of helium-4 ($Y_p$) and deuterium ($10^5D/H$), we obtained $Y_p=0.2588 \pm 0.0001$ and $10^5 D/H=2.88 \pm 0.04$, respectively \cite{Almaraz19_BDE}. These bounds represent an excess of 5 and 12 percent with respect to the $\Lambda$CDM constraints $Y_p=0.2467 \pm 0.0001$ and $10^5 D/H=2.59 \pm 0.04$, respectively. Although the difference is manifest, the constraints in BDE are still compatible with current astrophysical data \cite{Aver13_BBN,Izotov14_BBN,Riemer15_BBN,Zavarygin18_BBN,Cooke18_BBN}, considering the dispersion of the bounds\footnote{See, for example, table 5 of reference \cite{Zavarygin18_BBN}.} and the systematics affecting them \cite{Cyburt16_BBN}.

\begin{figure}[t!]\begin{center}\includegraphics[width=1\textwidth, height=0.3\textheight]{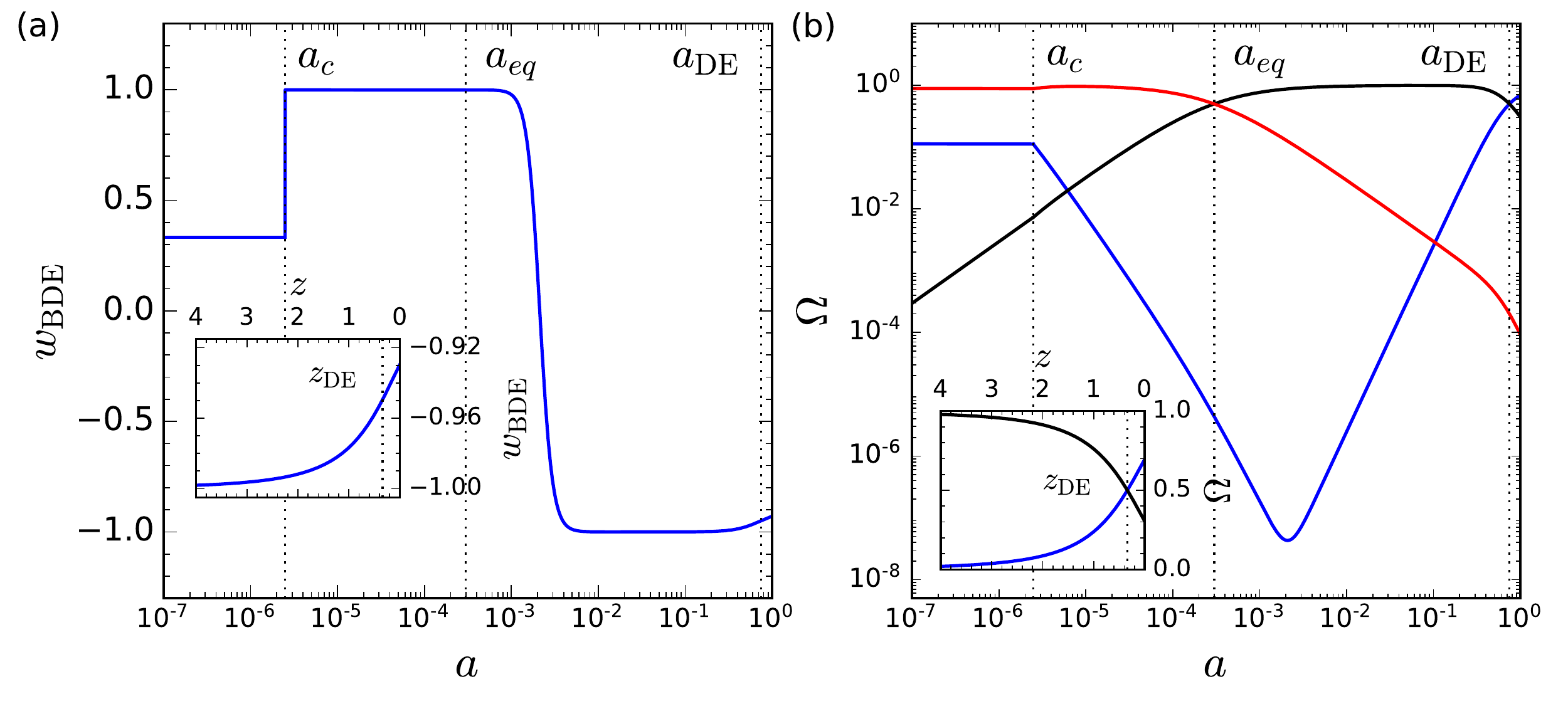}\caption{(a) Evolution of the Equation of State (EoS) of dark energy in the BDE model. (b) Density parameter of radiation (red), matter (black) and dark energy (blue) in BDE. Inner subplots show these quantities at late times as a function of the cosmological redshift $z$. Vertical dotted lines mark the condensation epoch of BDE ($a_c$), matter-radiation ($a_{eq}$) and matter-dark energy ($a_\mathrm{DE}$) equivalence, respectively.}\label{fig:bde_eos}\end{center} \end{figure}

We now describe the dynamics of DE in our model. Figure \ref{fig:bde_eos}a shows the evolution of the EoS over time. Before condensation, when the DG is present the EoS is simply $w_\textrm{BDE}=1/3$. When condensation occurs, the EoS leaps abruptly to $w_\textrm{BDE}\simeq 1$ and the scalar field behaves as a stiff fluid for a while. Next, shortly before recombination ($z_* \approx 1090$) the EoS drops to $w_\textrm{BDE}\simeq -1$ and now the scalar field mimics a cosmological constant $w_\Lambda = -1$. Finally, the EoS grows at late times reaching its present value $w_\textrm{BDE0} = -0.9294 \pm 0.0007$, which is shown in more detail in the inner subplot. Note how cosmological data tightly constrain the EoS in our model. Figure \ref{fig:bde_eos}b shows the density parameter $\Omega_i = \rho_i/\rho_\textrm{crit}$ (with $\rho_\textrm{crit}=3H^2/(8\pi G)$, see eqs. (\ref{eq:bde_Friedmann_DG}) and (\ref{eq:bde_Friedmann_phi}) below) of matter, radiation and DE. Shortly before condensation the DG amounts to $\Omega_\textrm{DG}=0.112$ the energy content of the universe. This is also the initial density parameter of BDE, since the energy of the DG is completely transferred to the scalar BDE mesons at $a_c$ \cite{Macorra03_BDE}. When the EoS leaps to $w_\textrm{BDE}\simeq 1$ the energy density of BDE redshifts as $\rho_\textrm{BDE} \propto a^{-6}$ and the scalar field dilutes more quickly than matter and radiation. This rapid dilution of BDE just after $a_c$ leaves and interesting imprint on the growth of matter perturbations in linear theory and one of the goals we pursue in this research is to find out how much of this signature is still present when nonlinear dynamics is taken into account. The scalar field is rapidly diluted and it remains subdominant for most of the history of the universe. When the EoS drops to $-1$ mimicking a cosmological constant, $\rho_\textrm{BDE}=const$ and the density parameter of BDE reaches its minimum value and then starts growing, since both matter and radiation are further diluted. Finally, $w_\textrm{BDE}$ departs from $-1$ at late times, the matter-DE equality epoch arrives at $z_\textrm{DE}=0.34$, and BDE becomes dominant with $\Omega_\textrm{BDE0}=0.696 \pm 0.007$ today.
\begin{figure}[t!]\begin{center}\includegraphics[width=1\textwidth, height=0.3\textheight]{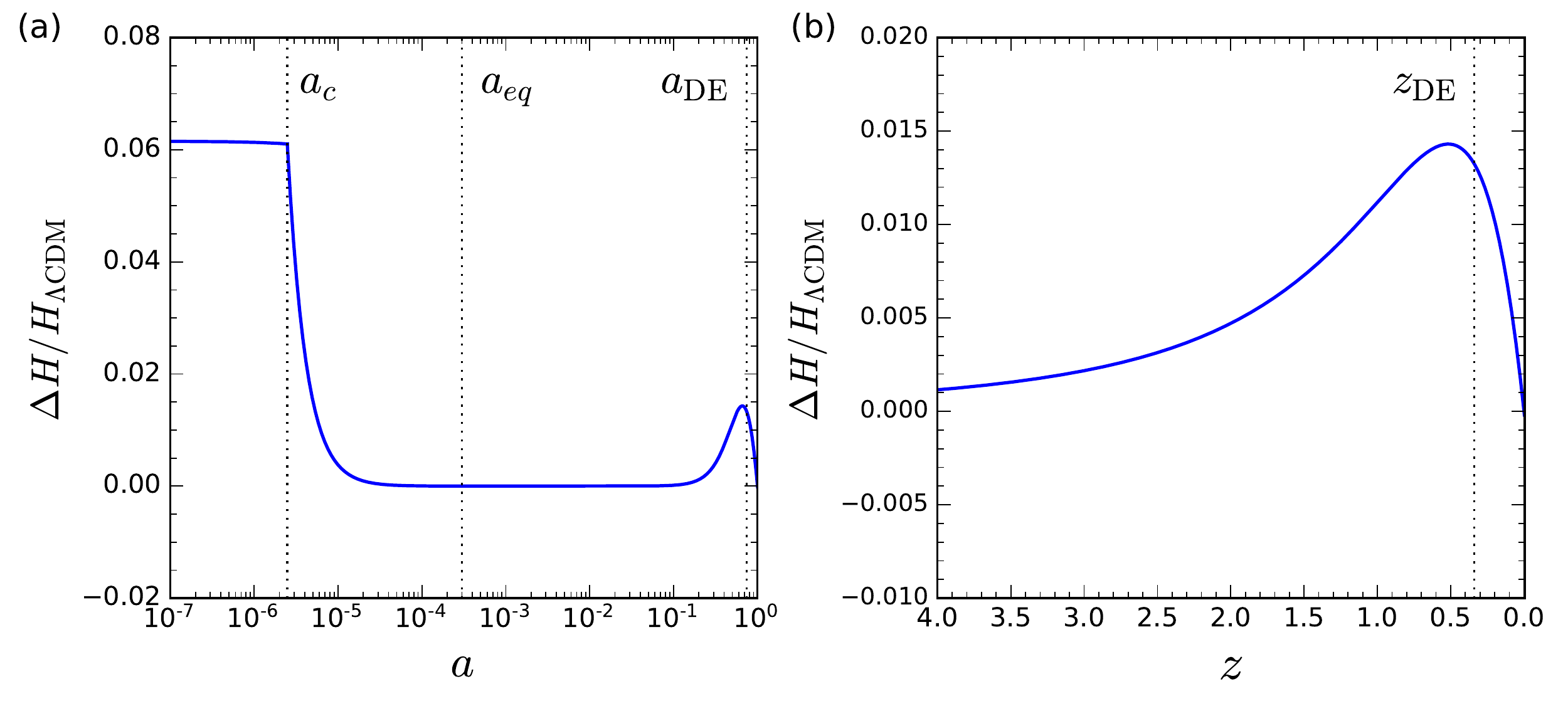}\caption{(a) Relative difference of the expansion rate ($H$) in BDE with respect to $\Lambda$CDM over time. In both models, the present amount of matter is $\Omega_m=0.305$ and the expansion rate today is $H_0=67.68 \textrm{ km s}^{-1}\textrm{Mpc}^{-1}$. Vertical dotted lines mark the same epochs as in figure \ref{fig:bde_eos}. (b) Relative difference of $H$ at late times as a function of the cosmological redshift $z$.}\label{fig:bde_hubble_rate}\end{center} \end{figure}

The immediate effect of this DE dynamics is reflected in the expansion rate of the universe. Before $a_c$ when the DG is present we have
\begin{equation}\label{eq:bde_Friedmann_DG} 
H^2 = \frac{8\pi G}{3}\left(\rho_{m}+\rho_{r}+\rho_\mathrm{DG}\right) \hspace*{0.5cm} \textrm{ for } a<a_c,
\end{equation}
where $\rho_m \propto a^{-3}$ and $\rho_r \propto a^{-4}$ are the energy densities of matter and SM radiation, respectively, and $\rho_\textrm{DG}=(2\Lambda_c^4/[1-w_\textrm{BDE}(a_c)])a^{-4}$ is the energy density of the DG, with $w_\textrm{BDE}(a_c)=1/3$. After condensation, the expansion rate is
\begin{equation} \label{eq:bde_Friedmann_phi} 
H^2=\frac{8\pi G}{3}(\rho_{m}+\rho_{r}+\rho_\textrm{BDE}) \hspace*{0.5cm} \textrm{ for } a\geqslant a_c,
\end{equation}
with $\rho_\textrm{BDE}$ given by the first expression of eq. (\ref{eq:bde_rhophi_Pphi}). The modification of the expansion rate leaves interesting imprints on observable quantities that can be used to constrain the model. A full discussion on how BDE fits observations and how these fits compare with those of $\Lambda$CDM can be found in \cite{Macorra18_BDE,Almaraz19_BDE}. Here we will focus instead on the phenomenological implications on structure formation that arise solely because of the different DE dynamics. Therefore, from now on we shall use the same set of cosmological parameters when we compare our results with the $\Lambda$CDM predictions. In this way, any difference we observe is due to the effects of DE. Figure \ref{fig:bde_hubble_rate}a shows the relative difference $(H-H_{\Lambda\textrm{CDM}})/H_{\Lambda\textrm{CDM}}$ with respect to a $\Lambda$CDM cosmology with $H^2_{\Lambda\textrm{CDM}}=\frac{8\pi G}{3}(\rho_m + \rho_r + \rho_\Lambda)$, and $\rho_\Lambda = \Lambda/(8\pi G)$. Since both models run with the same set of parameters, $H_0$ and $\Omega_m$ are the same, which automatically implies an equal amount of DE at present time (since $\rho_r$ depends only on the CMB temperature today). Initially, the DG enhances the expansion rate in BDE by about the $6\%$. However, once the particles of the DG condense into BDE the scalar field quickly dilutes leaving only matter and standard radiation. Since $\rho_\Lambda$ in $\Lambda$CDM is still negligible at these times, by the matter-radiation equality epoch ($a_{eq}$) arrives the expansion rate in both models is practically the same and therefore the relative difference is null. Later on, when DE becomes relevant, the expansion rate in BDE is once again larger as shown in figure \ref{fig:bde_hubble_rate}b in more detail. As we mentioned before, the density of DE and the expansion rate today is the same in both models. However, as we run the picture backwards $\rho_\textrm{BDE} \propto a^{-3(1+w_\textrm{BDE})} > \rho_\Lambda$, since $w_\mathrm{BDE}>-1$ and therefore the expansion rate in BDE is larger. The maximum deviation with respect to $\Lambda$CDM at late times only amounts to $1.4\%$ at $z\approx 0.5$, very close to $z_\textrm{DE}$. As we proceed to earlier times, $\rho_\textrm{BDE}$ is frozen to a nearly constant value, since $w_\textrm{BDE}$ approaches to $-1$. However, although the expansion rate is still larger in BDE, the difference with respect to $\Lambda$CDM gradually vanishes as the matter becomes dominant. The end result of all these processes is the bump in the plots.   

\section{Linear and quasilinear structure formation in BDE}\label{sec:bde_linear_quasilinear}
\subsection{Linear perturbation theory}\label{ssec:bde_linear_quasilinear_linear}
Large-scale structure formation is mainly driven by the dynamics of the CDM involving only gravitational physics. In the standard picture, structure formation proceeds from the collapse of initial small perturbations of CDM into dense knots called halos, which in turn cluster into filaments and sheets \cite{Cooray02_halo}. The resulting DM network serves as the skeleton around which ordinary matter accretes to form stars and galaxies. By observing these objects we expect to trace back the underlying structure and learn its properties. The effects of DE on this process come from the expansion rate and  DE perturbations. Unlike the cosmological constant, alternative models introduce DE inhomogeneities, which do have an impact on the evolution of radiation and matter perturbations \cite{Ma99_QCDM,Brax00_QCDM,Weller03_QCDM,Bean04_QCDM,Mehrabi15_DE}. In general terms, the approach to structure formation depends on the size of the density fluctuations ($\delta \equiv \delta \rho /\bar{\rho}$) around the homogeneous background ($\bar{\rho}$) and is roughly divided into three regimes: the linear, the quasilinear, and the nonlinear regime. 
\begin{figure}[t!]\begin{center}\includegraphics[width=1\textwidth, height=0.3\textheight]{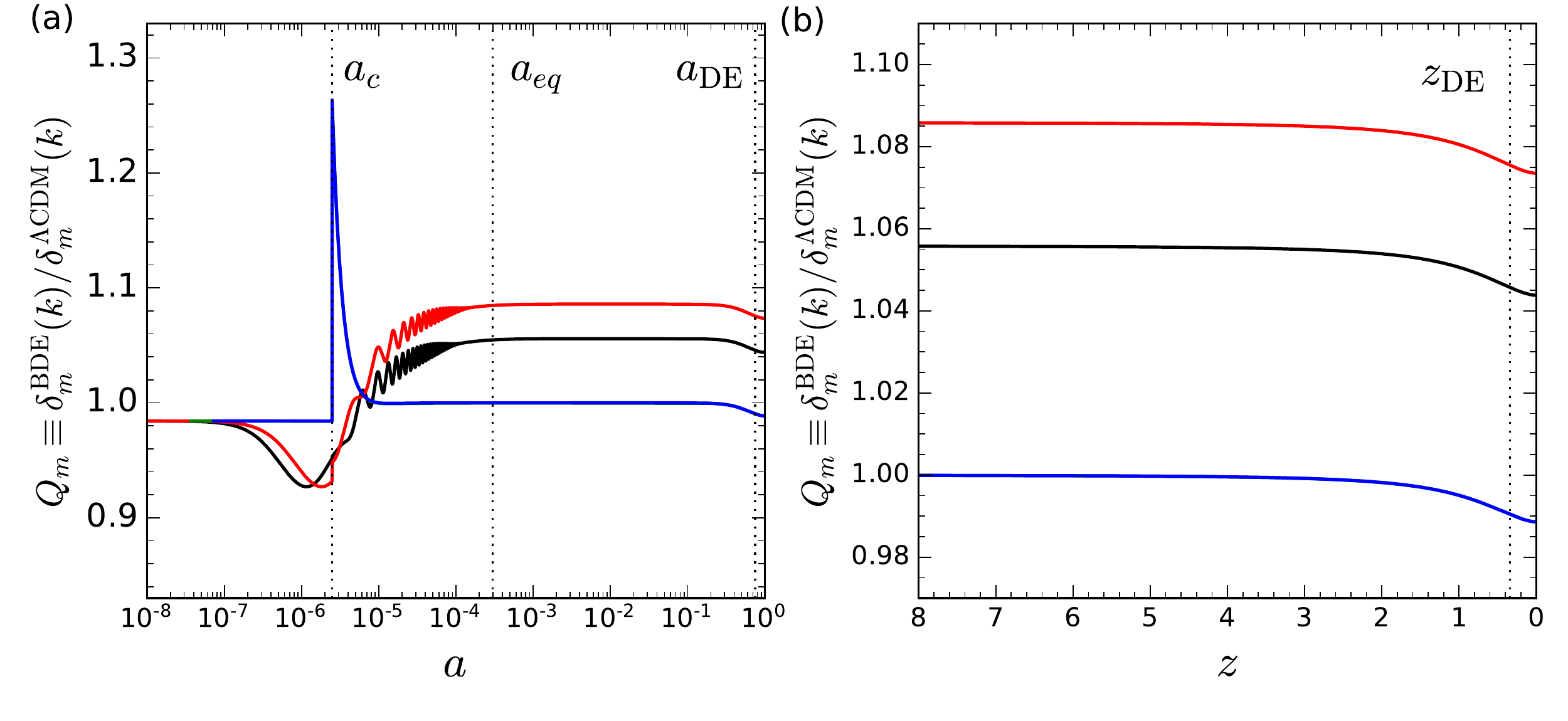}\caption{(a) Ratio of matter perturbations (in Newtonian gauge) between BDE and $\Lambda$CDM for modes $k [h\textrm{Mpc}^{-1}]=10$ (black), $6.37$ (red), $0.1$ (green) and $0.05$ (blue). The curve for $k=0.05 h\textrm{Mpc}^{-1}$ overlaps the curve for $k=0.1 h\textrm{Mpc}^{-1}$. The vertical dotted lines mark the same epochs as in figures \ref{fig:bde_eos} and \ref{fig:bde_hubble_rate}. (b) Ratio of matter perturbations at late times (as a function of the cosmological redshift, $z$) for the same modes.}\label{fig:bde_deltam}\end{center} \end{figure}

In the linear regime density fluctuations remain small enough to be accurately described by linear perturbation theory, which can be efficiently implemented in Boltzmann codes such \texttt{CAMB} \cite{Lewis00_CAMB}. We have studied the effects of BDE on the evolution of matter perturbations in the linear regime in \cite{Macorra18_BDE,Almaraz19_BDE}. Here we summarize our findings. Figure \ref{fig:bde_deltam}a shows the ratio $Q_m\equiv\delta_m^{\mathrm{BDE}}(k)/\delta_m^{\Lambda\mathrm{CDM}}(k)$ of matter perturbations in BDE and $\Lambda$CDM for different modes $k$ in the Newtonian gauge defined by the scalar potentials $\Psi$ and $\Phi$ through the line element $ds^2=-(1+2\Psi)dt^2+a^2(1-2\Phi)\delta_{ij}dx^idx^j$, with $\delta_{ij}$ the Kronecker delta \cite{MaBertschinger95_Cosmo}. Initially, $Q_m$ is constant since all the modes lie outside the horizon and therefore they do not evolve over time yet, as expected in this gauge. The initial suppression in BDE is due to the extra relativistic degrees of freedom of the DG affecting the initial overdensities $\delta_m$ through $\Psi$ \cite{Almaraz19_BDE,MaBertschinger95_Cosmo}. Perturbations start evolving after the horizon-entry epoch $a_h$ defined implicitly by $k=a_hH(a_h)$ for each mode. We see that small modes $k>k_c=a_cH(a_c)=1.37h\textrm{ Mpc}^{-1}$ crossing the horizon before $a_c$ evolve quite differently than large modes $k<k_c$ whose horizon-entry epoch occurs after that. 

Small modes $k>k_c$ are further suppressed in BDE since the entry epoch is delayed with respect to $\Lambda$CDM because of the DG. However, once they cross the horizon also in BDE, the growth rate is larger than in $\Lambda$CDM and therefore $Q_m$ increases leaving the troughs we see before $a_c$. Next, when condensation of BDE occurs the EoS leaps abruptly to $1$ and the energy density is rapidly diluted as $\rho_\mathrm{BDE}\propto a^{-6}$ (see figure \ref{fig:bde_eos}). Although the universe expands faster in BDE as shown in figure \ref{fig:bde_hubble_rate}a, the growth rate of matter perturbations is further enhanced since deceleration $\ddot{a}/a=-8\pi G (2\rho_r + 4\rho_\mathrm{BDE})/3$ is more efficient because of the extra (fading) term $4\rho_\mathrm{BDE}$, which is larger than $2\rho_r$ by 25\% at $a_c$. As a result, $Q_m$ is boosted above $1$ for these modes, reaching a maximum of $Q_m=1.085$ for $k=6.37 h\textrm{ Mpc}^{-1}$ before matter-radiation equivalence. After that, matter overdensities scale simply as $\delta_m \propto a$ during matter domination and $Q_m$ is constant again. Finally, when dark energy becomes dominant at late times the growth rate in BDE is smaller and $Q_m$ is uniformly suppressed for all modes by $-1.14\%$ the constant value it had during matter domination, as shown in detail in figure \ref{fig:bde_deltam}b. On the other hand, large modes $k<k_c$ are mainly affected by this late-time signature of the model. The transition from the DG to BDE leaves perturbations with the same amplitude $\delta_m \propto \Phi$ as in $\Lambda$CDM, since these modes are still outside the horizon and $\Phi$ is the same once the relativistic degrees of freedom of the DG vanish at $a_c$ \cite{MaBertschinger95_Cosmo}. The transient leap seen in the plot is nothing more than a gauge effect arising from the leap of the EoS on the synchronous potential $\eta$ used to convert matter overdensities to Newtonian gauge in $\texttt{CAMB}$.

Although density fluctuations are not gauge-invariant quantities, the resulting enhancement of matter perturbations is a real effect of BDE. We can see this by recalling how CDM overdensities transform between the synchronous and the Newtonian gauge \cite{MaBertschinger95_Cosmo},
\begin{equation}\label{eq:gauge_transf}
\delta_{cdm}(\mathrm{Syn})= \delta_{cdm}(\mathrm{New})-\frac{3aH}{k^2}\theta(\mathrm{New}),
\end{equation}
where $\theta(\mathrm{New})$ is the divergence of the velocity of CDM in the Newtonian gauge. As we can see, gauge effects are important on large scales, where $aH \gg k$. As times goes by and modes enter the horizon ($aH \ll k$) the second term in eq. (\ref{eq:gauge_transf}) becomes small, and the value of the overdensity coincides in both gauges leading to the same ratio $\delta_m^{\mathrm{BDE}}(k)/\delta_m^{\Lambda\mathrm{CDM}}(k)$ at late times.

\begin{figure}[t!]\begin{center}\includegraphics[width=0.57\textwidth, height=0.3\textheight]{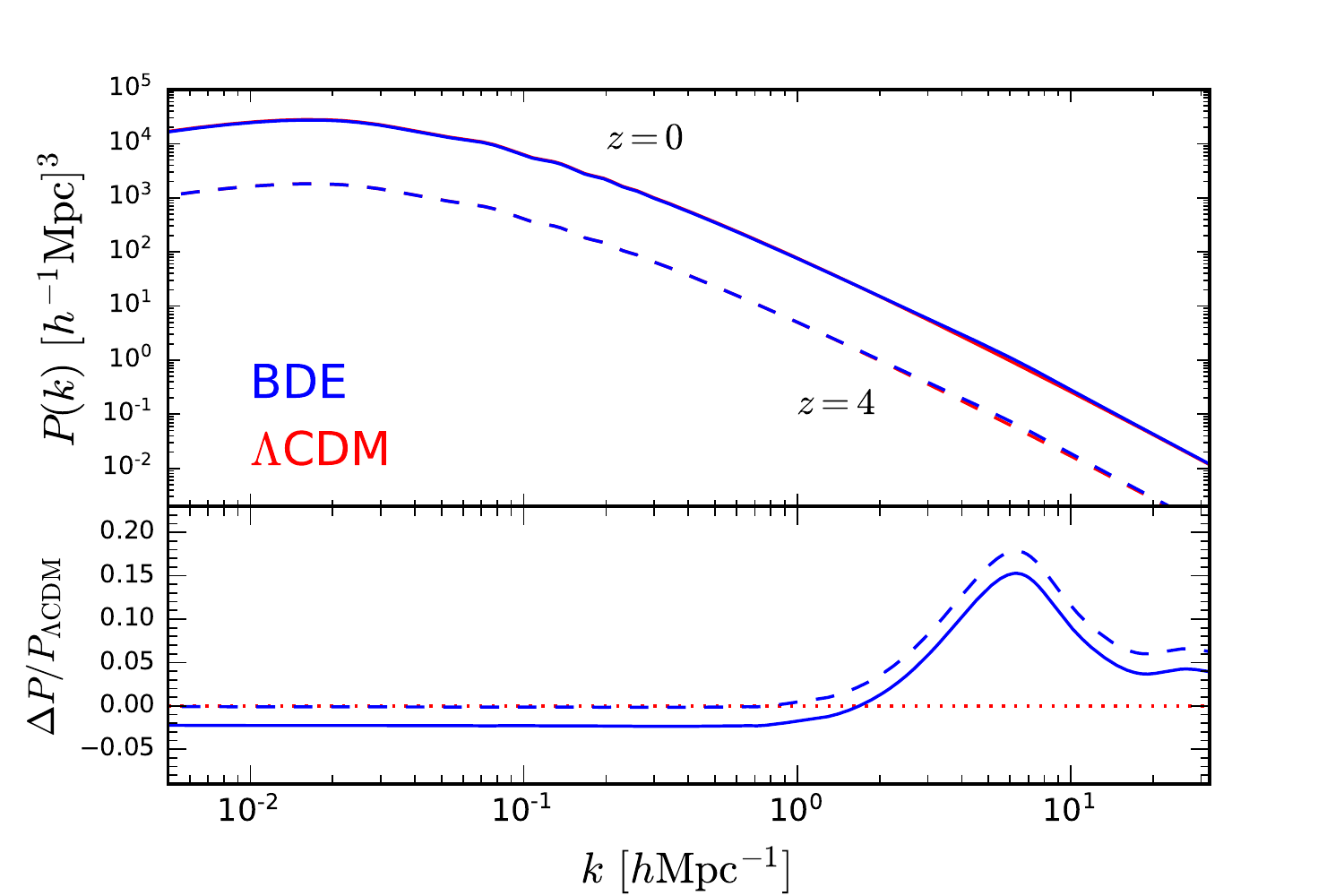}\caption{Matter power spectrum at $z=0$ (solid) and $z=4$ (dashed) for BDE (blue) and $\Lambda$CDM (red) in linear perturbation theory. The bottom panel shows the relative difference with respect to $\Lambda$CDM.}\label{fig:linear_power_spectrum}\end{center} \end{figure}

We see that BDE impacts the evolution of matter perturbations not only at late times as is expected, but there is also a distinctive imprint left by BDE condensation and rapid dilution. These two effects are manifest in the matter power spectrum in figure \ref{fig:linear_power_spectrum}, where the lower panel displays the relative difference with respect to $\Lambda$CDM. At $z=0$ the spectrum in BDE is generally suppressed by $2\%$ on all scales because of the late-time dynamics of DE, but this suppression is overwhelmed on small scales by the enhancement effect due to BDE rapid dilution at $a_c$, which leads to an excess of power of $15\%$ at $k\approx 6.37 h\textrm{ Mpc}^{-1}$. In fact, we can isolate these two effects by looking at the spectrum at earlier times, when DE is not dominant yet. For example, at $z=4$ matter perturbations on large scales have the same amplitude as in $\Lambda$CDM as shown in figure \ref{fig:bde_deltam}. At that time the spectrum is not suppressed yet and the only feature we see in figure \ref{fig:linear_power_spectrum} is the excess of power on small scales. 

It is important to bear in mind the physical context where these predictions hold. First of all, modes evolve uncoupled each other in linear theory, which means that there is no transfer of features between different scales. Secondly, the imprint left by BDE dilution on the matter spectrum is a relic effect of physical phenomena that took place in the early universe, almost 5 $e-$folds before matter-radiation equivalence. The small modes where this feature is imprinted have grown enough to enter the nonlinear regime, where new phenomena comes into play and linear perturbation theory is not valid anymore. Among the questions we address in this research is to determine how much of this signature remains when nonlinear phenomena are taken into account.

\subsection{The spherical collapse model in BDE}\label{ssec:bde_linear_quasilinear_quasilinear}
Before considering the dynamics of fully-nonlinear structures, it is very instructive to look at an intermediate regime where the transition from small perturbations to collapsing objects takes place. The spherical collapse model is the simplest approximation capturing some of the main features of this process. In the basic implementation of the model we consider a spherical overdense region embedded in a flat FLWR background. Initially, the overdense region has a radius $R_i$ and expands together with the background at a progressively decreasing rate until reaching a maximum size $R_\mathrm{max}$, where the expansion of the sphere halts and is subsequently reversed. According to the model, the contraction of the sphere after the turnaround epoch continues until the whole region collapses to a point $R_c=0$ and the density diverges to infinity. This unphysical state is not reached in nature, but the collapse of the sphere is also halted by a relaxation mechanism leaving the region in a final state of virial equilibrium with some finite size.

The evolution of a spherical overdense region with a top-hat density profile and radius $R$ is given by
\begin{equation} \label{eq:sc_radius}
\frac{\ddot{R}}{R}=-\frac{4\pi G}{3}(\rho + 3P),
\end{equation}
where $\rho=\bar{\rho}+\delta \rho$ and $P=\bar{P}+\delta P$ are the total energy density and pressure in the sphere. The formation of non-linear structures takes place at relatively late times, where the content of radiation in the universe can be neglected. Moreover, since DE in BDE is also very homogeneous on sub-horizon scales \cite{Almaraz19_BDE}, we may assume that density inhomogeneities are restricted to matter, and therefore  $\rho=\bar{\rho}_m(1+\delta_m)+\bar{\rho}_\mathrm{BDE}$ and $P=\bar{P}_\mathrm{BDE}$. Equation (\ref{eq:sc_radius}) is coupled with the Klein-Gordon (\ref{eq:Klein-Gordon}) and Friedmann (\ref{eq:bde_Friedmann_phi}) background equations as well as with a conservation equation for the total matter density within the sphere, $\dot{\rho}_m+3(\dot{R}/R)\rho_m=0$. These expressions can be combined \cite{Courtin11_HMF,Pace10_SC} into a single equation of motion for $\delta_m$ as
\begin{equation}\label{eq:sc_delta}
\ddot{\delta}_m+2H\dot{\delta}_m-4\pi G \bar{\rho}_m\delta_m=4\pi G \bar{\rho}_m\delta_m^2+\frac{4}{3}\frac{\dot{\delta}_m^2}{1+\delta_m},
\end{equation}
where the linear and non-linear terms are clearly identified on the left and the right hand side of the equation, respectively. The influence of DE is reflected in the friction term proportional to $H$, which is different from cosmology to cosmology. Initially $\delta_m$ is small enough to neglect quadratic terms and the system evolves linearly. However, as time goes by, these terms become relevant and the system enters the non-linear regime, where $\delta_m$ grows until diverging at a time of collapse $t_c$.

Solving eq. (\ref{eq:sc_delta}) allows us to get the linear extrapolation at $t_c$ of the critical initial overdensity that leads to collapse at that time ($\delta_c$). This quantity can be used to estimate the abundance of structures in the universe and therefore to search for possible imprints of DE \cite{Mainini03_SC,Pace10_SC}. In an Einstein-de Sitter cosmology ($\Omega_m=1$) it is found that $\delta_c^\mathrm{EdS}\approx 1.686$, with no dependence on the redshift \cite{Amendola10_DE}. This result is no longer valid when DE is present. Following \cite{Pace10_SC,Courtin11_HMF}, to get $\delta_c$ in BDE we first determine the initial condition for $\delta_m$ in eq. (\ref{eq:sc_delta}) leading to collapse at a given input redshift $z_c$. We choose the initial time $z_{ini}=300$ and set $\dot{\delta}_m(z_{ini})=0$, since the sphere starts expanding with the Hubble flow. The value of $\delta_m(z_{ini})$ can be efficiently determined through a binary search using trial values $\delta_m^\mathrm{min}(z_{ini})$ and $\delta_m^\mathrm{max}(z_{ini})$ whose redshifts of collapse encapsulate $z_c$ as $z_c^\mathrm{max}<z_c<z_c^\mathrm{min}$. The search stops when convergence below an accuracy threshold $\Delta z = 10^{-4}$ around $z_c$ is achieved. Once we have $\delta_m(z_{ini})$ and using $\dot{\delta}_m(z_{ini})=0$ we solve the linear version of eq. (\ref{eq:sc_delta}), where the right hand side is zero, and get $\delta_c$ as the value of the solution at $z_c$.

\begin{table}[t!] 
	\centering
	{\setlength{\extrarowheight}{2.5pt}
		\begin{tabular}{|l c c c c c|}\hline
			\multirow{2}{*}{Model} & $z_c=1$ & $z_c=0.8$ & $z_c=0.5$ & $z_c=0.2$ & $z_c=0$\\
			& $\delta_c$ & $\delta_c$ & $\delta_c$ & $\delta_c$ & $\delta_c$  \\\hline
			BDE & 1.686 & 1.684 & 1.682 & 1.679 & 1.675\\
			$\Lambda$CDM & 1.686 & 1.685 & 1.683 & 1.680 & 1.676 \\
			\hline
	\end{tabular}} 
	\caption{\label{tab:deltac}Linear extrapolation at $z_c$ of the critical initial overdensity that leads to collapse at that time in the spherical collapse model. Both in BDE and $\Lambda$CDM the present content of matter is $\Omega_m=0.305$ and the expansion rate today is $H_0=67.68 \textrm{ km s}^{-1}\textrm{Mpc}^{-1}$.}		
\end{table}

Table \ref{tab:deltac} shows the value of $\delta_c$ in BDE for different redshifts. We attach the corresponding results for standard $\Lambda$CDM with the same content of matter and expansion rate today. At large $z$ matter is dominant and $\delta_c$ tends to $\delta_c^\mathrm{EdS}$ irrespective of the dynamics of the DE, as expected. On the other hand, as DE becomes relevant at lower redshifts, the amount of Hubble friction is larger, which suppresses linear growth and leads to $\delta_c < \delta_c^\mathrm{EdS}$ both in BDE and $\Lambda$CDM. However, from figure \ref{fig:bde_hubble_rate}b we see that the amount of Hubble friction is larger in BDE and therefore $\delta_c$ is still smaller, as confirmed in table \ref{tab:deltac}. Nevertheless, the difference $\Delta \delta_c/\delta_c^{\Lambda\mathrm{CDM}} \lesssim 0.06 \%$ for $0\leqslant z \leqslant 1$ is small enough to be of some importance.

The abundance of structures in the universe is expressed by the halo mass function (HMF), which quantifies the distribution of the number density of halos per unit mass. From an experimental point of view, to know the HMF is important because it allows to constrain cosmological parameters from observations of galaxy clusters \cite{Allen11_LSS}. On the theoretical side, the HMF is an input ingredient for modeling galaxy formation \cite{Somerville99_LSS}. There are many estimations \cite{Murray13b_HMF} of the HMF available in the literature, all of which can be compactly written as
\begin{equation} \label{eq:quasilinear_hmf} 
\frac{dn}{d\ln M} = f(\sigma_R) \frac{\bar{\rho}_{m0}}{M}\frac{d\ln \sigma^{-1}_R}{d\ln M},
\end{equation}
where $dn/d\ln M$ is the differential number density of structures per logarithmic mass bin, $\bar{\rho}_{m0}$ is the background density of matter today, and $f(\sigma_R)$ is a function of the variance of the matter overdensity field smoothed on a comoving length scale $R$, which is given by
\begin{equation} \label{eq:quasilinear_sigma} 
\sigma^2_R(z)=\frac{1}{2\pi^2}\int_0^\infty k^3 P(k,z)|W_R(k)|^2d\ln k,
\end{equation}
where $P(k,z)$ is the linear matter power spectrum, and $W_R(k)$ is the Fourier transform of a top-hat spherical function of radius $R$ enclosing a mass $M=4/3\pi \bar{\rho}_{m0}R^3$. 

\begin{figure}[t!]\begin{center}\includegraphics[width=0.57\textwidth ]{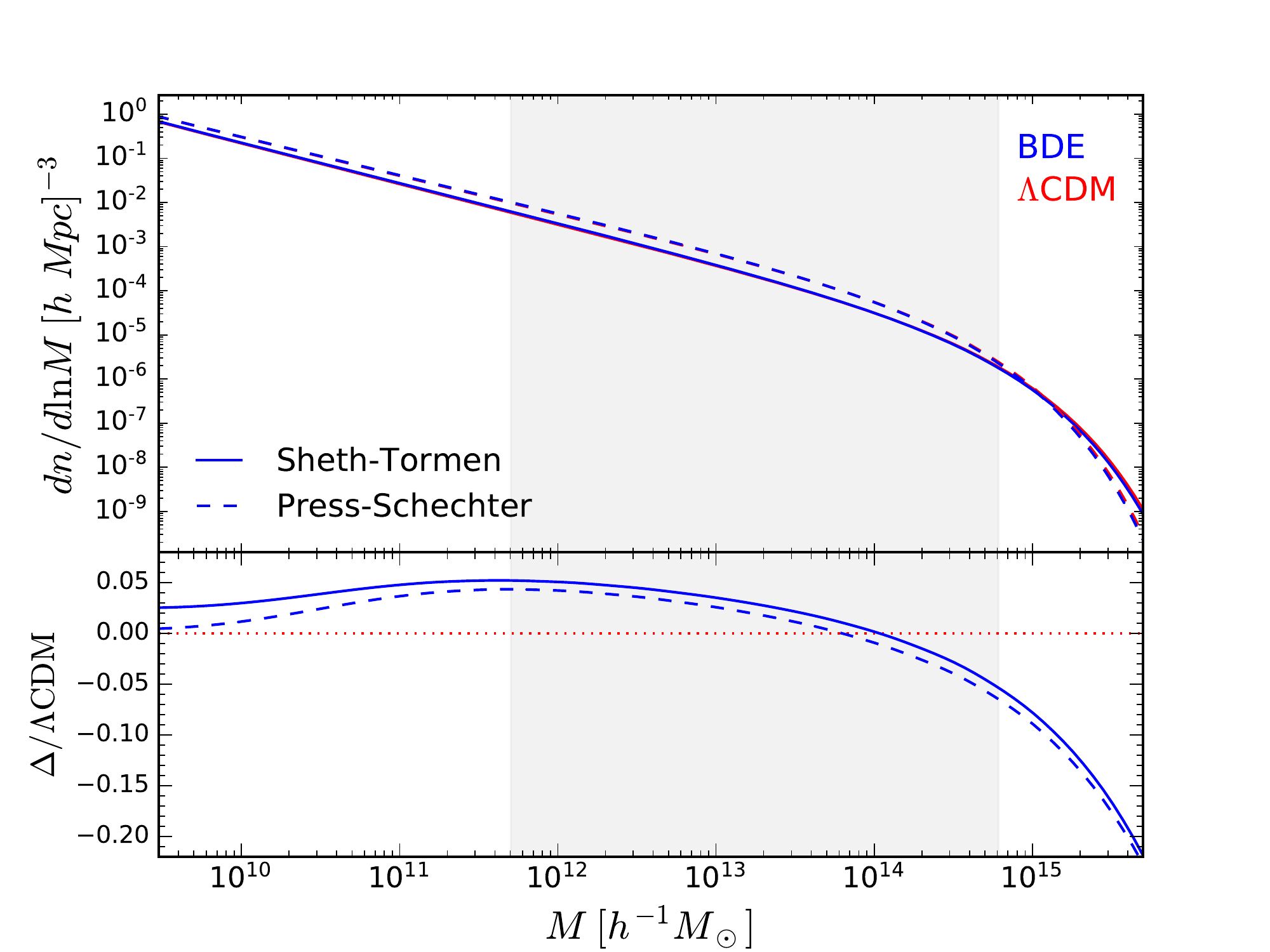}\caption{Mass function for BDE (blue) and $\Lambda$CDM (red) at $z=0$ using the Press-Schechter \cite{PressSchechter74_SF} and the Sheth-Tormen \cite{ShethTormen01_SF} formulas. The lower panel shows the relative difference with respect to $\Lambda$CDM. The shaded area marks the mass range probed by our N-body simulations (section \ref{ssec:bde_nonlinear_hmf}).}\label{fig:analytical_hmf}\end{center} \end{figure}

The explicit functional form of $f(\sigma_R)$ may be completely deduced from theoretical considerations as done in the original Press-Schechter theory \cite{PressSchechter74_SF}, where spherical collapse is assumed and the abundance of structures is estimated by counting the number of overdense regions above $\delta_c$ in a Gaussian random field. In this case, $f(\sigma_R)$ is given by
\begin{equation} \label{eq:quasilinear_press_schechter} 
f_\mathrm{PS} = \sqrt{\frac{2}{\pi}} \frac{\delta_c}{\sigma_R} \mathrm{exp}\left( -\frac{\delta_c^2}{2\sigma^2_R}\right)
\end{equation}
A more realistic approach considering ellipsoidal collapse instead of spherical collapse is proposed in \cite{ShethTormen01_SF},
\begin{equation} \label{eq:quasilinear_sheth_tormen} 
f_\mathrm{ST} = A\sqrt{\frac{2a}{\pi}} \left[ 1 + \left( \frac{\sigma^2_R}{a\delta_c^2} \right)^p\right] \frac{\delta_c}{\sigma_R} \mathrm{exp}\left( -\frac{a\delta_c^2}{2\sigma^2_R}\right),
\end{equation}
\renewcommand{\thefootnote}{\fnsymbol{footnote}}
where $A=0.3222$, $a=0.707$\footnote{Not to be confused with the cosmological scale factor.}, and $p=0.3$ are additional parameters whose value is set by calibrating with numerical simulations. These additional parameters are not completely independent, but they are related by a normalization condition \cite{Jenkins01_HMF,Bhattacharya11_HMF}
\begin{equation}\label{eq:st_condition}
\int_0^\infty f(\sigma) d\ln \sigma = 1,
\end{equation}
which means that all of the dark matter is clustered in halos.
The Press-Schechter and the Sheth-Tormen formulas are the only analytical distributions derived at present time \cite{Murray13a_HMF}. Other forms of $f(\sigma_R)$ are either extensions of these functions or empirical formulas designed to improve the fit to the HMF measured in numerical simulations. A thoroughly study of the adequacy of the different HMF to our simulations is beyond the scope of this research. Here we will limit our analysis to the Sheth-Tormen formula. However, even in this case the evidence gathered over the last years shows that if eq. (\ref{eq:quasilinear_sheth_tormen}) is to accurately fit simulations, there is a dependence on the redshift and cosmology that is necessary to take into account \cite{Crocce10_HMF,Courtin11_HMF,Bhattacharya11_HMF}. In practical terms this means that the parameters $A$, $a$, and $p$ have to be recalibrated according to the model and the epoch being considered. We delve into the effects of recalibrating these parameters in section \ref{ssec:bde_nonlinear_hmf}.

Figure \ref{fig:analytical_hmf} shows the predicted mass function for BDE and $\Lambda$CDM at $z=0$ using the Press-Schechter and the Sheth-Tormen formulas. We have used the corresponding value of $\delta_c$ of table \ref{tab:deltac} and the recalibrated values of $A$, $a$, and $p$ (see table \ref{tab:hmf_st_fit}). For the sake of reference, the shaded area shows the mass range spanned in our N-body simulations (see section \ref{ssec:bde_nonlinear_hmf}). Despite the systematic discrepancy between the Press-Schechter and the Sheth-Tormen predictions, the difference between BDE and $\Lambda$CDM is very similar. We see that there are less structures in BDE for $M\gtrsim 1\times 10^{14} h^{-1}M_\odot$, where we found that $\sigma_{\Lambda\mathrm{CDM}}>\sigma_\mathrm{BDE}$ and consequently the mass function is more suppressed by the exponential term in $f$. The opposite occurs for light structures, where the abundance is enhanced in BDE reaching an excess of $4\%$ at $M \approx 4.1\times 10^{11} h^{-1}M_\odot$. 

\section{Nonlinear structure formation in BDE}\label{sec:bde_nonlinear}
\subsection{N-body simulations setup}\label{ssec:bde_nonlinear_setup}
\begin{table}[t!] 
	\centering
	{\setlength{\extrarowheight}{2.5pt}
		\begin{tabular}{|l|c|}\hline
			Number of particles & $N_{part}=512^3$\\
			Box size  & $L_{box}=200 h^{-1}$ Mpc \\
			Mass of DM particles & $M_{part}=5.05 \times 10^{9} h^{-1}M_{\odot}$\\
			Initial redshift & $z_{ini}=49$\\
			Final redshift & $z_0=0$ \\
			Realizations & 5 \\\hline		
			Baryon density & $\Omega_bh^2=0.02252$\\
			Dark matter density & $\Omega_ch^2=0.1173$\\
			Dimensionless Hubble parameter & $h=0.6768$\\
			Spectral scalar index & $n_s=0.9774$\\
			Amplitude parameter & $A_s=2.367\times 10^{-9}$\\
			Optical depth at reionization & $\tau=0.117$\\
			\hline	 \end{tabular}}
\caption{\label{tab:results_setup} Technical specifications and cosmological parameters of the N-body simulations considered in this paper. The present amount of matter and DE is $\Omega_m=0.305$ and $\Omega_\mathrm{DE}=1-\Omega_m=0.695$, respectively. Given the basic cosmological parameters, the corresponding value of $\sigma_8$ at $z=0$ (see eq. (\ref{eq:quasilinear_sigma})) is $\sigma_8^{\mathrm{BDE}} = 0.854$ for BDE and $\sigma_8^{\Lambda\mathrm{CDM}} = 0.864$ for $\Lambda$CDM, respectively.}		
		\end{table}

In order to determine how BDE affects structure formation in the nonlinear regime, we prepared a suite of numerical simulations using the adaptive mesh refinement N-body code \texttt{RAMSES} \cite{Teyssier02_RAMSES}. This code allows the computation of high-resolution gravitational interactions by automatically refining the spatial grid in those regions where the number of particles exceeds some threshold. Since BDE does not appreciably cluster on sub-horizon scales \cite{Almaraz19_BDE},  we modeled DE as an homogeneous component in our simulations \cite{Baldi12_DEsimulations}, and thus we solely modified \texttt{RAMSES} to account for the corresponding background expansion  rate $H$, eq. (\ref{eq:bde_Friedmann_phi}). This can be efficiently done by interpolating from a precomputed table $(a,H)$ instead of solving the Klein-Gordon (\ref{eq:Klein-Gordon}) and Friedmann (\ref{eq:bde_Friedmann_phi}) equations during the run \cite{Dolag04_DEsimulations}. Table \ref{tab:results_setup} lists the setup of our simulations and the cosmological parameters we used in our analysis. This is the same set of cosmological parameters we have been using in our previous discussions. The corresponding density parameter of matter and DE today are $\Omega_m=0.305$ and $\Omega_\mathrm{DE}=1-\Omega_m=0.695$, respectively. We also prepared a suite of N-body simulations for $\Lambda$CDM using the same configuration. As we mentioned before, we shall use the same cosmological parameters to isolate the effects of the dynamics of DE on the quantities we study. We ran five realizations in each model, varying the seed for the initial conditions, which were accordingly prepared using the second-order Lagrangian perturbation code \texttt{2LPTic} \cite{Crocce06b_2lptic}. All simulations ran from $z_{ini}=49$ up to the present epoch $z_0=0$. For BDE we modified \texttt{2LPTic} so that the program just takes as input the full linear power spectrum from \texttt{CAMB} at $z_{ini}$ and returns the corresponding positions and velocities of the DM particles at that time. Later on, we measured the spectrum at $z_{ini}$ in our simulations and verified that the initial linear spectrum of \texttt{CAMB} is consistently retrieved. 

\subsection{Matter power spectrum}\label{ssec:bde_nonlinear_mps}
\begin{figure}[t!]\begin{center}\includegraphics[width=1.\textwidth, height=0.32\textheight]{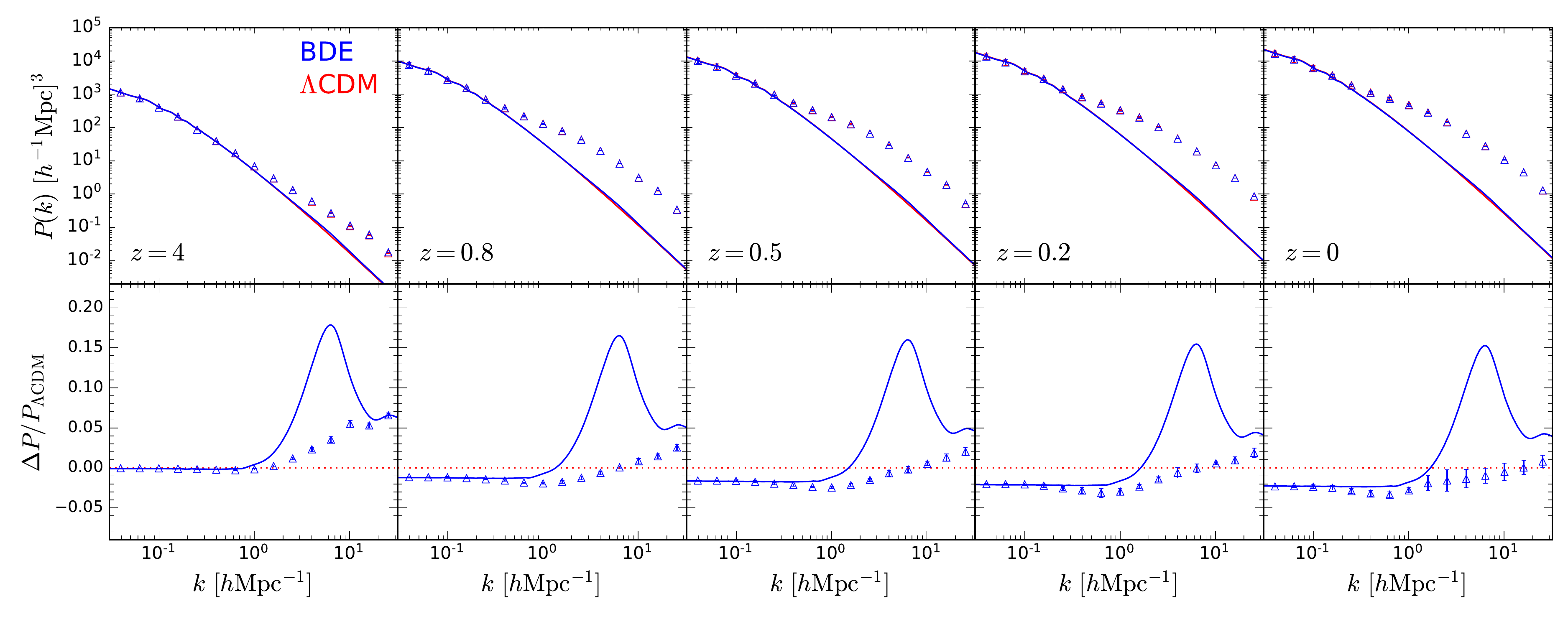}\caption{Matter power spectrum in BDE (blue) and $\Lambda$CDM (red) at different redshifts. In the top panel the symbols show the average of the nonlinear spectra measured in our simulations across the five realizations, while the error bars indicate the standard deviation. The solid curves show the predictions from the linear theory. The lower panels display the relative difference of BDE with respect to $\Lambda$CDM of the nonlinear power spectrum (symbols) and the linear theory (solid).}\label{fig:results_power_spectrum}\end{center} \end{figure}

We measure the matter power spectrum in our simulations using the public code \texttt{POWMES} \cite{Colombi09_POWMES} with a grid size of $N_g=2N_{part}^{1/3}=1024$ to achieve a high resolution. This allows us to measure the spectrum up to $k = 32h\textrm{ Mpc}^{-1}$ large enough to probe any excess of power on small scales, as predicted by linear theory. Figure \ref{fig:results_power_spectrum} shows our results at different redshifts. The solid lines in the top panels correspond to the spectra obtained from linear theory ($P_\textrm{l}$) using \texttt{CAMB}, while the symbols show the binned spectra measured in our simulations ($P_\textrm{nl}$), and the error bars indicate the standard deviation across the five realizations for each model. The lower panels show the relative difference with respect to $\Lambda$CDM in linear theory (solid) and in our simulations (symbols). 

On large scales $k\lesssim 0.1h\textrm{ Mpc}^{-1}$, where linear theory is valid and the spectrum is mainly determined by the spatial distribution of the halos \cite{Cooray02_halo,Ma07_mps}, the results of our simulations agree with linear theory, as expected. At high redshifts there is no difference with respect to $\Lambda$CDM, but as time goes by the spectrum in BDE is gradually suppressed from $-1\%$ at $z=0.8$ to $-2\%$ today. This feature is just the late-time suppression effect due to the distinct DE dynamics in BDE, which affects all the modes in the same way, as we previously discussed in section \ref{ssec:bde_linear_quasilinear_linear}. In the intermediary regime $0.1h\textrm{ Mpc}^{-1} \lesssim k\lesssim 1h\textrm{ Mpc}^{-1}$ nonlinear corrections start to come into play especially at late times, where collapsing structures have evolved so much since they broke away from the background expansion. However, the departures from $\Lambda$CDM in the simulations are essentially the same as those the linear theory predicts, except for the slight suppression near $k= 1h\textrm{ Mpc}^{-1}$, which at $z=0$ drives further the difference between BDE and $\Lambda$CDM to $-3\%$ around $k\approx 0.6h\textrm{ Mpc}^{-1}$. 

\begin{figure}[t!]\begin{center}\includegraphics[width=0.57\textwidth, height=0.3\textheight]{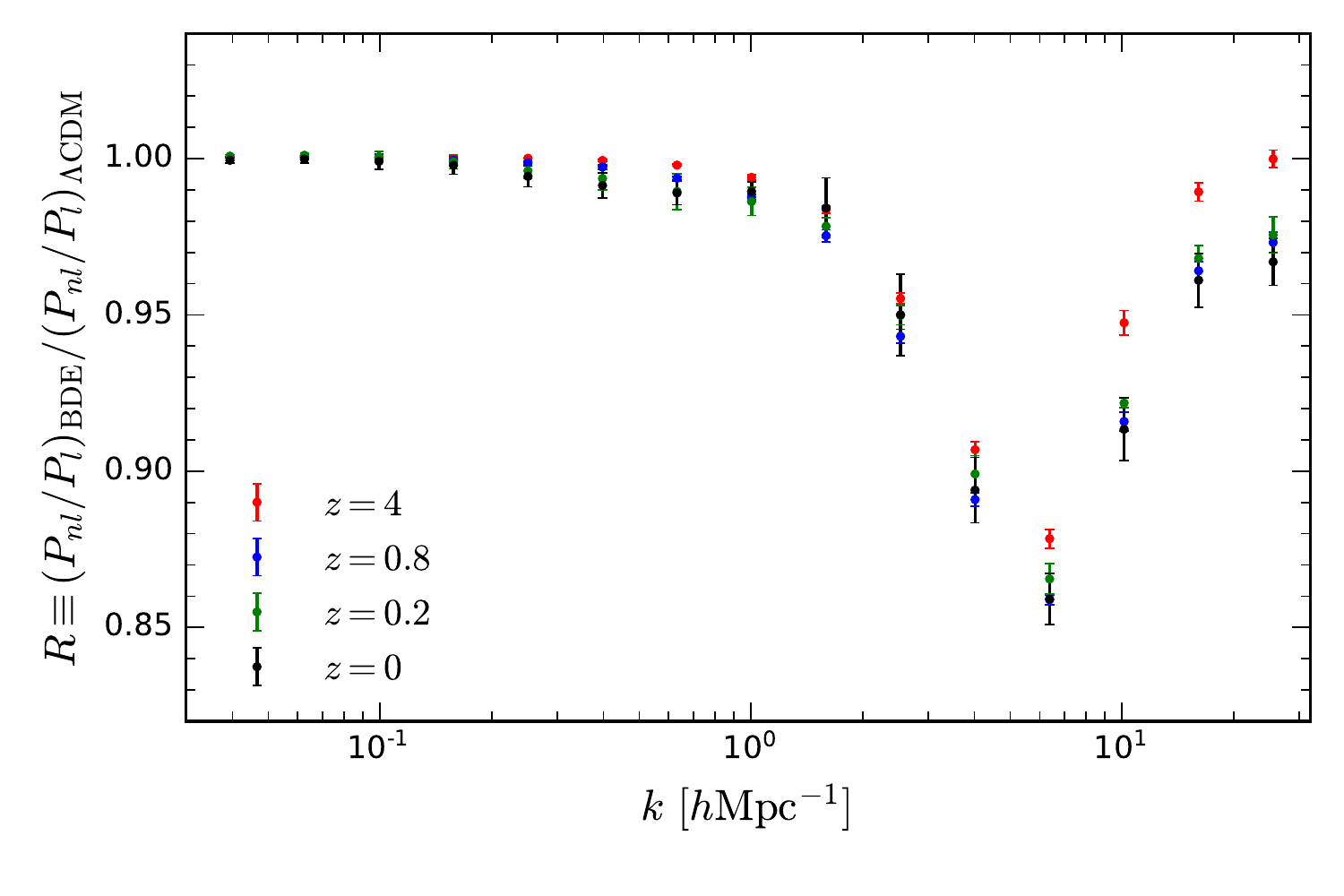}\caption{Ratio of the nonlinear power spectrum ($P_\mathrm{nl}$) to the predictions of the linear theory ($P_\mathrm{l}$) in BDE and $\Lambda$CDM at different redshifts. The symbols show the average across the five realizations and the error bars indicate the standard deviation.}\label{fig:results_rratio}\end{center} \end{figure}

It is interesting to see what happens on small scales $k\gtrsim 1h\textrm{ Mpc}^{-1}$. We recall that in this regime linear theory predicts more power in BDE as a consequence of the modification of the expansion rate of the universe when the scalar field is rapidly diluted after $a_c$. The enhancement of matter perturbations affects the modes $k\gtrsim k_c = 1.37h\textrm{ Mpc}^{-1}$ crossing the horizon before $a_c$, leaving a maximum deviation with respect to $\Lambda$CDM at $k_\textrm{max}\approx 6.37h\textrm{ Mpc}^{-1}$, as seen in the plots of the bottom panels of figure \ref{fig:results_power_spectrum}. Note that the position of the peaks remains the same and it is only the height that is decreasing because of the mode-independent late-time suppression effect. When we take into account nonlinear gravity interactions in our N-body simulations, power is transferred from large to small scales \cite{Ma07_mps} and the excess of power in BDE is washed out leaving only a weak trace below $1\%$ at $z=0$ on the smallest scales we were able to probe. The dilution of the peak proceeds long before the linear late-time suppression effect becomes relevant. For example, although there is still more power in BDE at $z=4$, the difference with respect to $\Lambda$CDM in the simulations is already very distinct from linear theory. However, once linear suppression comes into play it also affects the residuals in the simulations, as seen by the uniform downward drift of the markers between $z=0.8$ and $z=0.2$. Finally, at late times the difference in the nonlinear spectra decouples from the linear suppression effect, as shown by the mild extra flattening of the markers at $z=0$ on the smallest scales. We can compare how much nonlinear effects are present in each model through the ratio \cite{Alimi10_QCDM}
\begin{equation}\label{eq:results_R}
R=\frac{(P_\textrm{nl}/P_\textrm{l})_\textrm{BDE}}{(P_\textrm{nl}/P_\textrm{l})_{\Lambda\textrm{CDM}}},
\end{equation}  where each term $P_\textrm{nl}/P_\textrm{l}$ accounts for the difference between the spectrum measured in the simulations and the spectrum obtained from linear theory. Figure \ref{fig:results_rratio} shows this ratio at different times. On large scales nonlinear corrections are small in both models and therefore $R\approx 1$ as expected. On smaller scales, where nonlinear dynamics becomes dominant $R<1$, which means that $(P_\textrm{nl}/P_\textrm{l})_{\Lambda\textrm{CDM}}>(P_\textrm{nl}/P_\textrm{l})_\textrm{BDE}$ and consequently the nonlinear corrections are more acute in $\Lambda$CDM, thus diluting the gained power in BDE since $a_c$. Moreover, nonlinear effects in $\Lambda$CDM are stronger at late times as shown in the plot, where $R(z=4)>R(z\ll 1)$ for the smallest scales. 

\subsection{Halo mass function}\label{ssec:bde_nonlinear_hmf}
\begin{figure}[t!]\begin{center}\includegraphics[width=1.\textwidth, height=0.32\textheight]{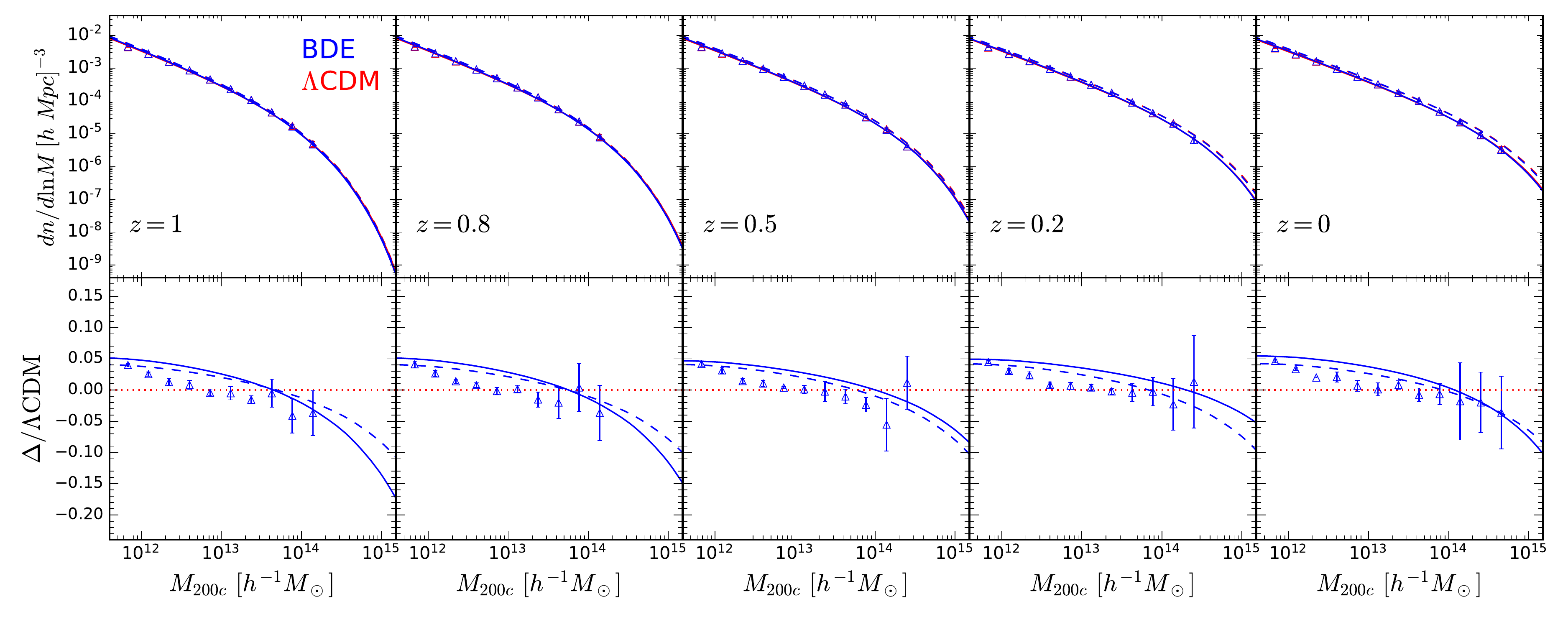}\caption{Halo mass function (HMF) in BDE (blue) and $\Lambda$CDM (red) at different redshifts. In the top panels the symbols show the average of the HMF measured in our simulations across the five realizations, while the error bars indicate the standard deviation. We have selected halos consisting of at least 100 DM particles, $M_{200c}>100M_{part}=5.05\times 10^{11}h^{-1}M_\odot$, with subhalos filtered out. The solid curves show the predictions from the Sheth-Tormen formula \cite{ShethTormen01_SF} after recalibrating the parameters $A$, $a$, and $p$ in eq. (\ref{eq:quasilinear_sheth_tormen})---see table \ref{tab:hmf_st_fit}--- and using the corresponding values of the density collapse $\delta_c$ of table \ref{tab:deltac}, while the dashed lines show the distributions with the standard values $A=0.3222$, $a=0.707$, and $p=0.3$. The lower panels display the relative difference of BDE with respect to $\Lambda$CDM of the measured HMF in our simulations (symbols) and the Sheth-Tormen formula (solid and dashed lines).}\label{fig:results_HMF}\end{center} \end{figure}

We built the halo catalogs in our simulations using the halo finder \texttt{ROCKSTAR} \cite{Behroozi12_ROCKSTAR}. This code identifies dark matter halos and substructures by implementing an extended version in the 6D phase space of the Friends-of-Friends (FOF) algorithm, where overdense regions are identified by grouping particles according to some linking length. The halo mass is defined as the mass contained within a sphere of virial radius $R_{\Delta_c}$ whose mean density is $\Delta_c = 200$ times the critical density $\bar{\rho}_c$ of the universe \cite{White01_halomass}
\begin{equation}\label{eq:m200c_definition}
M_{200c}=\frac{4\pi}{3}\Delta_c\bar{\rho}_cR_{\Delta_c}^3
\end{equation}
To estimate the halo mass function we have considered parent halos consisting at least of 100 DM particles ($M_{200c}>100M_{part}$), which yields a resolution in mass of $5.05\times  10^{11}h^{-1}M_\odot$, and we also filtered subhalos out. Figure \ref{fig:results_HMF} shows our results. The symbols in the top panels display the average across the five simulations, while the error bars indicate the standard deviation, and the lower panels show the relative difference with respect to $\Lambda$CDM. We get accurate measurements of the halo mass function in our simulations for masses between $\sim 10^{12} h^{-1}M_\odot$ and $10^{14}h^{-1}M_\odot$ corresponding to galactic-sized and small cluster-sized halos, respectively. In all redshifts the halo mass function is larger in BDE in the low-mass end of the plots by 4\%. However, the difference with respect to $\Lambda$CDM decreases with the mass and eventually there are fewer heavy halos in BDE, although the large error bars due to the small number of structures in the high-mass end $M_{200c}\gtrsim 10^{14}h^{-1}M_\odot$ don't allow us to draw further quantitative conclusions. Interestingly, the location of the crossing point separating these two regions in the residuals shifts slightly to the right as we can see by comparing the snapshots at $z=1$ and $z=0$. The BDE model has a stronger clustering power on small scales since the initial time, which means that more small halos form in this regime. On the other hand, large heavier structures form more slowly in BDE for the same reason as $P(k)$ is suppressed at small $k$, namely, structure formation on large scales is affected strongly by the expansion history of the universe. 

\begin{table}[t!] 
	\centering
	{\setlength{\extrarowheight}{2.5pt}
		\begin{tabular}{|l | c c c | c c c|}\hline
			\multirow{2}{*}{Redshift} & \multicolumn{3}{c|}{BDE} & \multicolumn{3}{c|}{$\Lambda$CDM}\\
			& $A$ & $a$ & $p$ & $A$ & $a$ & $p$\\\hline
			$z=1$   & 0.290 & 0.706 & 0.330 & 0.287 & 0.700 & 0.333\\
			$z=0.8$ & 0.285 & 0.695 & 0.335 & 0.282 & 0.690 & 0.337\\
			$z=0.5$ & 0.282 & 0.718 & 0.337 & 0.281 & 0.720 & 0.338\\
			$z=0.2$ & 0.277 & 0.753 & 0.342 & 0.274 & 0.759 & 0.344\\
			$z=0$   & 0.270 & 0.785 & 0.347 & 0.267 & 0.780 & 0.350\\			
			\hline
	\end{tabular}} 
	\caption{\label{tab:hmf_st_fit}Recalibrated values of the parameters $A$, $a$, and $p$ of the Sheth-Tormen formula, eq. (\ref{eq:quasilinear_sheth_tormen}) for different redshifts. In each case, we used the corresponding value of $\delta_c$ listed in table \ref{tab:deltac}. Since the amplitude $A$ is fixed by the normalization condition eq. (\ref{eq:st_condition}), only $a$ and $p$ were freely varied in the fitting process.}		
\end{table}

We compared the halo abundances measured in our simulations with the predictions of the Sheth-Tormen formula by using the proper values of $\delta_c$ listed in table \ref{tab:deltac} and recalibrating the parameters $a$ and $p$ of eq. (\ref{eq:quasilinear_sheth_tormen}). This is done by fitting $a$ and $p$ to minimize the quantity $\sum [(dn/d\ln M)_\mathrm{ST}/(dn/d\ln M)_\mathrm{sim} -1]^2$, which accounts for the difference between the predictions of the Sheth-Tormen distribution $(dn/d\ln M)_\mathrm{ST}$ and the halo mass function measured in the simulations $(dn/d\ln M)_\mathrm{sim}$. The sum runs over the mass bins and the amplitude $A$ is inferred from the normalization condition eq. (\ref{eq:st_condition}). Table \ref{tab:hmf_st_fit} lists the recalibrated values of $A$, $a$, and $p$. Clearly, there is a moderate variation of these parameters over the time in each model (e.g., $\Delta A = 7.4\%$ from $z=0$ to $z=1$ in BDE). However, the differences between $\Lambda$CDM and BDE at a given redshift are only about $1\%$. The solid lines in figure \ref{fig:results_HMF} show the corresponding distributions, while the dashed lines are the distributions with the standard values $A=0.3222$, $a=0.707$, and $p=0.3$. The Sheth-Tormen formula describes the halo abundances in the simulations with a precision $\leqslant 15\%$ after recalibrating. This is particularly important at late times, where the formula with the standard values overpredicts the number of structures in the high-mass end, as seen in the top panels. At early times, recalibration does not represent a significant improvement. However, although the Sheth-Tormen formula captures the excess of small structures and the deficit of heavy objects in BDE, the residuals with respect to $\Lambda$CDM in the range $M_{200c}\sim 10^{12}-10^{13} h^{-1}M_\odot$ are a little bit overpredicted. In view of these results, we conclude that the Sheth-Tormen formula can be used to parametrize the abundances of structures in BDE, but this parametrization is still of limited utility to meet the needs of high-precision cosmology.

\subsection{Halo concentration parameter}\label{ssec:bde_nonlinear_concentration}
\begin{figure}[t!]\begin{center}\includegraphics[width=1.\textwidth, height=0.32\textheight]{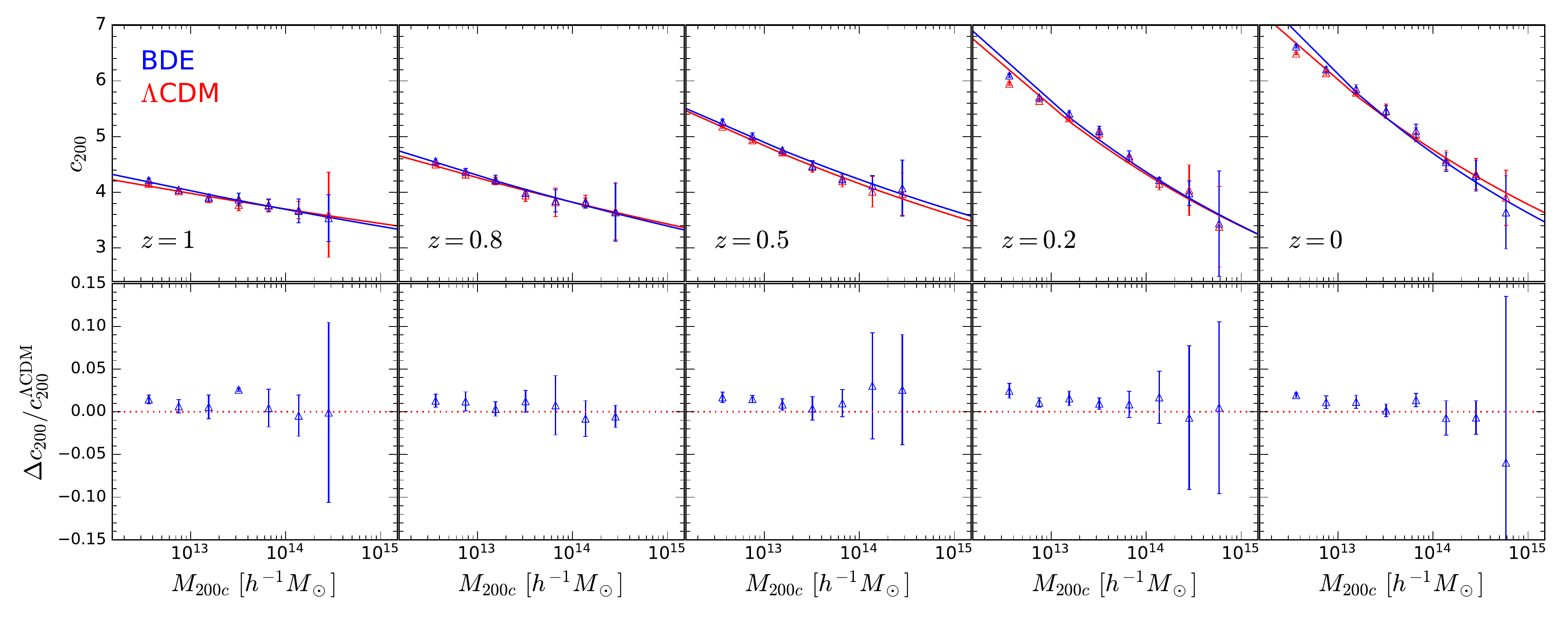}\caption{Halo concentration parameter in BDE (blue) and $\Lambda$CDM (red) at different redshifts. In the top panels the symbols show the average of the halo concentration parameter measured in our simulations across the five realizations, while the error bars indicate the standard deviation. We have selected massive halos consisting of at least 500 DM particles, $M_{200c}>500M_{part}=2.52\times 10^{12}h^{-1}M_\odot$ with subhalos filtered out. The solid curves are the best-fitting curves of the power-law function defined in eq. (\ref{eq:c_vs_M_fit}) with parameters given in table \ref{tab:results_c200_vs_M_parameters}. The lower panels display the relative difference of BDE with respect to $\Lambda$CDM of the halo concentration parameter measured in our simulations.}\label{fig:results_concentration}\end{center} \end{figure}

We measured the halo concentration parameter $c_{200}$ in our catalogs by computing the ratio
\begin{equation}
c_{200}=\frac{R_{200}}{r_s},
\end{equation} 
where $R_{200}$ is the virial radius given in eq. (\ref{eq:m200c_definition}), and $r_s$ is the characteristic radius obtained by fitting the halo to a Navarro-Frenk-White profile \cite{Behroozi12_ROCKSTAR,Prada12_concentration,NFW97_halos}. Figure \ref{fig:results_concentration} shows our results. We have selected from our catalogs massive halos with at least 500 DM particles ($M_{200c}>500M_{part}$) to balance the compromise between resolution and halo statistics in our simulations. In the top panel the symbols correspond to the mean values across the five realizations, while the error bars indicate the standard deviation. The solid lines show the fit to these data by the power-law function
\begin{equation}\label{eq:c_vs_M_fit}
c_{200}(z,M)=\alpha(z)\left( \frac{M}{10^{12}h^{-1}M_\odot} \right)^{\beta(z)},
\end{equation}
where the amplitude $\alpha$ and the slope $\beta$ depend on the redshift. We compile the best-fit values of these parameters in table \ref{tab:results_c200_vs_M_parameters}. The lower panel shows the relative difference of BDE with respect to $\Lambda$CDM in our simulations. We see that the concentration-mass relation in BDE has the same features found in $\Lambda$CDM \cite{Bullock01_concentration,Munoz11_concentration,Prada12_concentration}, to wit: i) the relation flattens for increasing $z$ as reflected in the progressively smaller values of the slope $\beta$ at late times in table \ref{tab:results_c200_vs_M_parameters}, ii) the concentration of halos of fixed mass increases with time, as shown by the increasing values of the amplitude $\alpha$, and iii) the concentration of halos at a given redshift decreases with the mass, as determined by the negative sign of $\beta$. The power-law function of eq. (\ref{eq:c_vs_M_fit}) fits well the concentration measured in our simulations, except in the low-mass end at $z\gtrsim 0.2$, where the function overpredicts the numerical results. However, we recall that the halo concentration-mass relation measured in other simulations exhibits more complex features to be captured by a simple fitting function \cite{Zhao09_concentration,Klypin11_halos,Prada12_concentration,Diemer15_concentration}, such as a flattening and an upturn for large masses, as well as a positive slope for large redshifts, which were beyond the reach of our simulations and require a more robust analysis.  

As far as the difference between BDE and $\Lambda$CDM is concerned, we don't find any substantial departure from $\Lambda$CDM, save a hint of more concentration in BDE by $\lesssim 1\%$ for $M_{200}\lesssim 10^{13} h^{-1}M_\odot$. In any case, these results show the low sensitivity of the halo concentration parameter to $H$, which suggests that clustering inside the halos is decoupled from the general expansion once they form.

\begin{table}[t!] 
	\centering
	{\setlength{\extrarowheight}{2.5pt}
		\begin{tabular}{|l c c c c c c c c c c|}\hline
			\multirow{2}{*}{Model} & \multicolumn{2}{c}{$z=1$} & \multicolumn{2}{c}{$z=0.8$}& \multicolumn{2}{c}{$z=0.5$} & \multicolumn{2}{c}{$z=0.2$} & \multicolumn{2}{c|}{$z=0$}\\
			& $\alpha$ & $\beta$ & $\alpha$ & $\beta$ & $\alpha$ & $\beta$ & $\alpha$ & $\beta$ & $\alpha$ & $\beta$ \\\hline
			\footnotesize{BDE} & \footnotesize{4.387} & \footnotesize{-0.037} & \footnotesize{4.841} & \footnotesize{-0.051} & \footnotesize{5.644} & \footnotesize{-0.063} & \footnotesize{7.204} & \footnotesize{-0.109} & \footnotesize{7.882} & \footnotesize{-0.112} \\
			\footnotesize{$\Lambda$CDM} & \footnotesize{4.279} & \footnotesize{-0.032} & \footnotesize{4.743} & \footnotesize{-0.047} & \footnotesize{5.610} & \footnotesize{-0.065} & \footnotesize{7.049} & \footnotesize{-0.106} & \footnotesize{7.540} & \footnotesize{-0.100} \\
			\hline
			 \end{tabular}} 
\caption{\label{tab:results_c200_vs_M_parameters}Best-fitting parameters of the power-law relation of eq. (\ref{eq:c_vs_M_fit}) to the halo concentration parameter measured in our simulations at different redshifts. These parameters were computed by performing a non-linear least squares fit as implemented in the \texttt{scipy} library of the \texttt{Python} package.}		
		\end{table}

\section{Summary and Conclusions}\label{conclusions}
In this paper we studied cosmic structure formation in the Bound Dark Energy (BDE) model. BDE is an alternative quintessence theory, where the scalar field describing DE is explained at a fundamental level by physics beyond the Standard Model. In BDE we introduce a hidden group of elemental light particles that are weakly coupled at high energies. These particles condense into composite states when a critical energy scale $\Lambda_c$ is reached and the gauge coupling of the hidden group becomes strong. DE is represented by the lightest formed state, which corresponds to a scalar meson particle described by a canonical scalar field $\phi$ with an IPL potential $V(\phi)=\Lambda_c^{4+2/3}\phi^{-2/3}$. The dynamics of DE and its cosmological implications differ from other quintessence scenarios, such as the Ratra-Peebles potential. Particularly, in BDE the expansion rate of the universe is affected not only at late times (as expected), but also in the early universe soon after condensation occurs. Interestingly, this leads to an enhancement (with respect to $\Lambda$CDM) of matter perturbations on small scales in linear perturbation theory.

The main issue we addressed in this paper was to investigate the impact of BDE on nonlinear structure formation through N-body simulations. Here we focused on the phenomenology of the model rather than analyzing how it fits cosmological observations. In order to identify the differences arising from the distinct DE dynamics, we compare our results with $\Lambda$CDM simulations with the same setup and cosmological parameters. Our results show that nonlinear gravitational interactions remove any trace of the enhancement predicted by linear theory from the late-time matter power spectrum. This is because nonlinear corrections on small scales are more pronounced in $\Lambda$CDM, thus compensating the initial gained power in BDE after condensation. However, it is still possible to observe remnants of this signature at redshifts within the reach of surveys. For example, at $z=4$ the BDE spectrum has more power than $\Lambda$CDM by $5\%$ on scales $k\gtrsim 10 h\textrm{ Mpc}^{-1}$. On the other hand, the spectrum in BDE is gradually suppressed on large scales as DE becomes dominant. At present time, the suppression amounts to $2\%$.

The halo mass function measured in our simulations shows an excess of small halos ($M_{200c}\sim 10^{12} h^{-1}M_\odot$) in BDE followed by a gradual suppression of heavy structures ($M_{200c}\gtrsim 10^{14}h^{-1}M_\odot$). These results suggest that nonlinear clustering proceeds more efficiently in BDE on small scales, while the formation of large heavier structures is delayed because of the general expansion. The Sheth-Tormen fitting formula provides a fair estimation of the halo mass function in BDE capturing up to some limited degree the difference with respect to $\Lambda$CDM. The halo concentration parameter measured in BDE follows the same behavior than standard $\Lambda$CDM and the concentration-mass relation is well fitted by a power-law relation. However, we did not find substantial differences with respect to $\Lambda$CDM, which suggests that clustering inside halos is decoupled from the general expansion once the halos form.

In view of all these results, we conclude that BDE and $\Lambda$CDM are strongly degenerated in the nonlinear regime of structure formation. However, we stress that our analysis was limited to the case when both models run with the same set of cosmological parameters. It might occur that when we compare realistic scenarios, the small differences we found here become more pronounced. So far, the main source for discriminating BDE and $\Lambda$CDM comes from BAO measurements and BBN nucleosynthesis \cite{Macorra18_BDE,Almaraz19_BDE}. There is still the intermediate regime of voids where we can look at. We leave this possibility for a future work.

\acknowledgments
E. A. is supported by the National Council of Science and Technology (CONACYT) [grant number 2019-000012-01EXTV-00354]. We acknowledge the financial support from projects CONACYT Fronteras 281, PASPA-DGAPA UNAM, and PAPIIT-DGAPA Project IN103518. E.A. thanks the hospitality of the Institute for Computational Cosmology (ICC) at Durham University for carrying out this research, Lydia Heck and Alastair Basden for technical support, and C\'esar Hern\'andez and Octavio Valenzuela for useful comments and discussions. This work used the DiRAC Data Centric system at Durham University, operated by the Institute for Computational Cosmology on behalf of the STFC DiRAC HPC Facility (www.dirac.ac.uk). This equipment was funded by BIS National E-infrastructure capital grant ST/K00042X/1, STFC capital grants ST/H008519/1 and ST/K00087X/1, STFC DiRAC Operations grant ST/K003267/1 and Durham University. DiRAC is part of the National E-Infrastructure.








\bibliographystyle{naturemag}
\bibliography{nonlinear_bde_jcap} 

\end{document}